\documentclass[a4paper,oneside,10pt]{article}%
\usepackage{amsmath}
\usepackage{amsfonts}
\usepackage{amssymb}
\usepackage{graphicx}
\usepackage[square,numbers,sort&compress]{natbib}%
\providecommand{\U}[1]{\protect\rule{.1in}{.1in}}

\newtheorem{theorem}{Theorem}[section]

\newtheorem{definition}[theorem]{Definition}
\newtheorem{assumption}[theorem]{Assumption}
\newtheorem{example}[theorem]{Example}

\newtheorem{lemma}[theorem]{Lemma}

\newtheorem{remark}[theorem]{Remark}

\numberwithin{equation}{section}

\begin{document}

\title{Recursive utility maximization under partial information}
\author{Shaolin Ji \thanks{Institute for Financial Studies, Shandong University, Jinan
250100, China and Institute of Mathematics, Shandong University, Jinan 250100,
China, Email: jsl@sdu.edu.cn.  }
\and Xiaomin Shi\thanks{Corresponding author. Institute for Financial Studies,
Shandong University, Jinan 250100, China, Email: shixm@mail.sdu.edu.cn.}}
\date{}
\maketitle

\textbf{Abstract}. This paper concerns the recursive utility maximization
problem under partial information. We first transform our problem under
partial information into the one under full information. When the generator of
the recursive utility is concave, we adopt the variational formulation of the
recursive utility which leads to a stochastic game problem and a
characterization of the saddle point of the game is obtained. Then, we study
the K-ignorance case and explicit saddle points of several examples are
obtained. At last, when the generator of the recursive utility is smooth, we
employ the terminal perturbation method to characterize the optimal terminal wealth.



\addcontentsline{toc}{section}{\hspace*{1.8em}Abstract}

\section{Introduction}

In this paper, we study the problem of an agent who invests in a financial
market so as to maximize the recursive utility of his terminal wealth $X(T)$
on finite time interval $[0,T]$, while the recursive utility is characterized
by the initial value $Y(0)$ of the following Backward Stochastic Differential
Equation (BSDE for short)
\begin{equation}
Y(t)=u(X(T))+\int_{t}^{T}f(s,Y(s),Z(s))ds-\int_{t}^{T}Z(s)d\widehat{W}%
(s).\label{intro-bsde}%
\end{equation}
The market consists of a riskless asset and $d$ risky assets, the latter being
driven by a d-dimensional Brownian motion. And the investor has access only to
the history of interest rates and prices of risky assets, while the
appreciation rate and the driving Brownian motion are not directly observed.
That is, the filtration generated by the Brownian motion could not be used
when the investor chooses his portfolios. This is quite practical in a real
financial market. So we are interested in this so called recursive utility
maximization problem under partial information.

In the full information case, the problem of maximizing the expected utility
of terminal wealth is well understood in a complete or constrained financial
market \cite{CK}, \cite{KLS}. In an incomplete multiple-priors model, Quenez
\cite{Qu} studied the problem of maximization of utility of terminal wealth in
which the asset prices are semimartingales. Schied \cite{Sc} studied the
robust utility maximization problem in a complete market under the existence
of a \textquotedblleft least favorable measure". As for the recursive utility
optimization, El Karoui et al\cite{EPQ} studied the optimization of recursive
utilities when the generator of BSDE is smooth. Epstein and Ji \cite{EJ1},
\cite{EJ} formulated a model of recursive utility that captures the
decision-maker's concern with ambiguity about both the drift and ambiguity and
studied the recursive utility optimization under G-framework. But all the
above works do not accommodate partial information.

In the partial information case, Lakner \cite{La} generalized the martingale
method to expected utility maximization problem, see also Pham \cite{Ph}. Cvitanic et al \cite{CLQZ}
maximized the recursive utiluty under partial information. But the generator
$f$ in Cvitanic et al \cite{CLQZ} doesn't depend on $z$. Miao \cite{Mi}
studied a special case of recursive multiple-priors utility maximization
problem under partial information in which the appreciation rate is assumed to
be an $F_{0}$-measurable, unobserved random variable with known distribution.
Actually, they studied the problem under Bayesian framework and did not give
the explicit solutions.

In this paper, we first transform our portfolio selection problem under
partial information into a one under full information in which the unknown
appreciation rate is replaced by its filter estimate and the Brownian motion
is replaced by the innovation process. Then, a backward formulation of the
problem under full information is built in which instead of the portfolio
process, the terminal wealth is regarded as the control variable. This
backward formulation is based on the existence and uniqueness theorem of BSDE
and was introduced in \cite{EPQ} and \cite{JZ1}.

When the generator $f$ of (\ref{intro-bsde}) is concave, we adopt the
variational formulation of the recursive utility which leads to a stochastic
game problem. Inspired by the convexity duality method developed in Cvitanic
and Karatzas \cite{CK1}, we turn the primal \textquotedblleft sup-inf" problem
to a dual minimization problem over a set of discounting factors and
equivalent probability measures. A characterization of the saddle point of
this game is obtained in this paper. Furthermore, the explicit saddle points
for several classical examples are worked out.

When the generator $f$ of the BSDE is smooth, we apply the terminal
perturbation method developed in Ji and Zhou \cite{JZ} and Ji and Peng
\cite{JP} to characterize the optimal terminal wealth of the investor. Once
the optimal terminal wealth is obtained, the determination of the optimal
portfolio process is a martingale representation problem which we do not
involve in this paper.

The rest of this paper is organized as follows. In section 2, we formulate the
recursive utility maximization problem under partial information, reduce the
original problem to a problem under full information and give the backward
formulation. The case of non-smooth generator is tackled in section 3. In
section 4, we specialize in K-ignorance model and give explicit saddle points
of several examples. In section 5, we characterize the optimal wealth when the
generator $f$ is smooth.
\section{The problem of recursive utility maximization under partial
observation}
\subsection{Classical formulation of the problem}
We consider a financial market consisting of a riskless asset whose price
process is assumed for simplicity to be equal to one, and d risky securities
(the stocks) whose prices are stochastic processes $S_{i}(t), i=0,1,...,d$
governed by the following SDEs:%

\begin{align}
\label{stock}dS_{i}(t)=S_{i}(t)\Big(\mu_{i}(t)dt+\sum\limits_{j=1}^{d}%
\sigma_{ij}(t)dW_{j}(t)\Big), i=1,...,d,
\end{align}
\newline where $W(\cdot)=(W_{1}(\cdot),...,W_{d}(\cdot))^{\prime}$ is a
standard d-dimensional Brownian motion defined on a filtered complete
probability space $(\Omega,\mathcal{F}, \{\mathcal{F}_{t}\}_{t\geq0}, P)$.
$\mu^{\prime}=\{\mu^{\prime}(t)=(\mu_{1}(t),...,\mu_{d}(t)), t\in[0,T]\}$ is
the appreciation rate of the stocks which is $\mathcal{F}_{t}$-adapted,
bounded, and the $d\times d$ matrix $\sigma(t)=(\sigma_{ij}(t))_{1\leq i,j\leq
d}$ is the disperse rate of the stocks. Here and throughout the paper
$^{\prime}$ denotes the transpose operator.


The asset prices are assumed to be continuously observed by the investors in
this market, in other words, the information available to the investors is
represented by $\mathbb{G}=\{\mathcal{G}_{t}\}_{t\geq0}$, which is the
P-augmentation of the filtration generated by the price processes
$\sigma(S(u);0\leq u\leq t)$. The matrix disperse coefficient $\sigma(t)$ is
assumed invertible, bounded uniformly and $\exists\varepsilon>0,
\ \rho^{\prime}\sigma(t)\sigma^{\prime}(t)\rho\geq\varepsilon||\rho||^2, \ \forall\rho\in\mathbb{R}^{d},
\ t\in[0,T], a.s.$. In fact, $\sigma(t)$ can be obtained from the quadratic
variation of the price process. So we assume w.l.o.g. that $\sigma(t)$ is
$\mathcal{G}_{t}$-adapted. However, the appreciation rate $\mu^{\prime
}(t):=(\mu_{1}(t),...,\mu_{d}(t))$ is not observable for the investors.

A small investor whose actions cannot affect the market prices can decide at
time $t\in[0,T]$ what amount $\pi_{i}(t)$ of his wealth to invest in the $i$th
stock, $i=1,...,d$. Of course, his decision can only be based on the available
information $\{\mathcal{G}_{t}\}_{t=0}^{T}$, i.e., the processes $\pi^{\prime
}(\cdot)=(\pi_{1}(\cdot),...,\pi_{d}(\cdot)):[0,T]\times\Omega\rightarrow
\mathbb{R}^{d}$ are $\{\mathcal{G}_{t}\}_{t=0}^{T}$ progressively measurable
and satisfy $E\int_{0}^{T} ||\pi(t)||^{2}dt<\infty.$

Then the wealth process $X(\cdot)\equiv X^{x,\pi}(\cdot)$ of a self-financing
investor who is endowed with initial wealth $x>0$ satisfies the following
stochastic differential equation:%

\begin{align}
\label{wealth1}dX(t)  &  = \sum\limits_{i=1}^{d}\pi_{i}(t)\frac{dS_{i}%
(t)}{S_{i}(t)} \ \nonumber\\
&  = \pi^{\prime}(t)\mu(t) dt+\pi^{\prime}(t)\sigma(t) dW(t).
\end{align}

Because the only information available to the investor is $\mathbb{G}$, we
could not use the Brownian motion $W$ to define the recursive utility. As we
will show in the following, there exists a Brownian motion $\widehat{W}$ under
P in the filtered measurable space $(\Omega,\mathbb{G})$ which is often
referred to as an innovation process. The recursive utility process
$Y(t)\equiv Y^{x,\pi}(t)$ of the investor is defined by the following backward
stochastic differential equation:
\begin{equation}
Y(t)=u(X(T))+\int_{t}^{T}f(s,Y(s),Z(s))ds-\int_{t}^{T}Z(s)d\widehat{W}(s),
\label{BSDE}%
\end{equation}
where $f$ and $u$ are functions satisfying the following assumptions.

\begin{assumption}
\label{assf1} (A1) $f: \Omega\times[0,T]\times\mathbb{R}\times\mathbb{R}%
^{d}\rightarrow\mathbb{R}$ is $\mathbb{G}$-progressively measurable for any
$(y,z)\in\mathbb{R}\times\mathbb{R}^{d}$.\newline(A2) There exists a constant
$C\geq0$ such that
\[
\big| f(t,y_{1},z_{1})-f(t,y_{2},z_{2})\big|\leq C(\big|y_{1}-y_{2}%
\big|+\big|z_{1}-z_{2}\big|),\ \forall(t, \omega, y_{1}, y_{2}, z_{1},
z_{2})\in\Omega\times[0,T]\times\mathbb{R}\times\mathbb{R}\times\mathbb{R}%
^{d}\times\mathbb{R}^{d}.
\]
(A3) $f(t,\cdot,\cdot)$ is continuous about t and $E\int_{0}^{T}%
f^{2}(t,0,0)dt<+\infty.$
\end{assumption}

\begin{assumption}
\label{assu1} $u:\mathbb{R}^{+}\rightarrow\mathbb{R}$ is continuously
differentiable and satisfies linear growth condition.
\end{assumption}

\begin{remark}
Equation (\ref{BSDE}) is not a standard BSDE because in general $\mathbb{G}$
is strictly larger than the augmented filtration of the $(P,\mathbb{G}%
)$-Brownian motion $\widehat{W}$.
\end{remark}

We introduce the following spaces:
\[%
\begin{array}
[c]{l}%
L^{2}(\Omega,\mathcal{G}_{T},P;\mathbb{R}):=\Big\{\xi:\Omega\rightarrow
\mathbb{R}\big|\xi\mbox { is }\mathcal{G}_{T}\mbox{-measurable, and }E|\xi
|^{2}<\infty\Big\},\\
M_{\mathbb{G}}^{2}(0,T;\mathbb{R}^{d}):=\Big\{\phi:[0,T]\times\Omega
\rightarrow\mathbb{R}^{d}\big|(\phi_{t})_{0\leq t\leq T}\mbox{ is }\mathbb{G}%
\mbox{-progressively measurable process,}\\
\mbox{  \ \ \ \  and }||\phi||^{2}=E\int_{0}^{T}|\phi(t)|^{2}dt<\infty
\Big\},\\
S_{\mathbb{G}}^{2}(0,T;\mathbb{R}):=\Big\{\phi:[0,T]\times\Omega
\rightarrow\mathbb{R}\big|(\phi_{t})_{0\leq t\leq T}\mbox{ is }\mathbb{G}%
\mbox{-progressively measurable process,}\\
\mbox{  \ \ \ \  and }||\phi||_{S}^{2}=E[\sup\limits_{0\leq t\leq T}%
|\phi(t)|^{2}]<\infty\Big\}.
\end{array}
\]

For notational simplicity, we will often write $L_{\mathcal{G}_{T}}^{2}$,
$M_{\mathbb{G}}^{2}$ and $S_{\mathbb{G}}^{2}$ instead of $L^{2}(\Omega
,\mathcal{G}_{T},P;\mathbb{R})$, $M_{\mathbb{G}}^{2}(0,T;\mathbb{R}^{d})$ and
$S_{\mathbb{G}}^{2}(0,T;\mathbb{R})$ respectively.

We will show in the next subsection that under Assumption \ref{assf1}, for any
$\xi\in L_{\mathcal{G}_{T}}^{2}$, the BSDE (\ref{BSDE}) has a unique solution
$(Y(\cdot),Z(\cdot))\in S_{\mathbb{G}}^{2}\times M_{\mathbb{G}}^{2}$. Then for
each $\pi\in M_{\mathbb{G}}^{2},\ X(T)\in L_{\mathcal{G}_{T}}^{2}$, and
Assumption \ref{assu1} ensures that the variable $u(X(T))\in L_{\mathcal{G}%
_{T}}^{2}$. Thus, under Assumptions \ref{assf1} and \ref{assu1}, the recursive
utility process associated with this terminal value is well defined.

Given an utility function satisfying Assumption \ref{assu1} and initial
endowment $x$, the recursive utility maximization problem with bankruptcy
prohibition is formulated as: the investor chooses a portfolio strategy so as
to
\begin{equation}
\mathrm{Maximize}\ \ \ Y^{x,\pi}(0),\label{optm}\\
s.t.%
\begin{cases}
X(t)\geq0,\ \ t\in\lbrack0,T],\ \ a.s.,\\
\pi(\cdot)\in M_{\mathbb{G}}^{2},\\
(X(\cdot),\pi(\cdot))\ \ \ \mathrm{satisfies\ \ \ Eq.}(\ref{wealth1}),\\
(Y(\cdot),Z(\cdot))\ \ \ \mathrm{satisfies\ \ \ Eq.}(\ref{BSDE}),
\end{cases}
\end{equation}
where $X(t)\geq0$ means that no-bankruptcy is required.

\begin{definition}
A portfolio $\pi(\cdot)$ is said to be admissible if $\pi(\cdot)\in
M_{\mathbb{G}}^{2}$ and the corresponding wealth processes $X(t)\geq
0,\ t\in\lbrack0,T],a.s.$.
\end{definition}

Given initial wealth $x>0$, denote by $\mathcal{\overline{A}}(x)$ the set of
investor's feasible portfolio strategies, that is
\[
{\mathcal{\overline{A}}}(x)=\Big\{\pi:\pi\in M_{\mathbb{G}}^{2},\ X^{x,\pi
}(t)\geq0,\ dP\otimes dt\ a.s.\Big\}.
\]

\subsection{Reduction to a problem under full information}

Define the risk premium process $\eta(t)=\sigma(t)^{-1}\mu(t)$. Because we
have assumed the process $\mu(\cdot),\sigma(\cdot)$ are uniformly bounded, the
process
\[
L(t):=\exp(-\int_{0}^{t}\eta^{\prime}(s)\mathrm{d}W(s)-\frac{1}{2}\int_{0}%
^{t}|\eta(s)|^{2}ds)
\]
is a $(P,\mathbb{F})$ martingale. So a probability measure $\widetilde{P}$ is
defined by%

\[
\widetilde{P}(A)=E[L(T)I_{A}], \ \forall A\in\mathcal{F}_{T}, \ \text{where}%
\ \frac{\mathrm{d}\widetilde{P}}{\mathrm{d}P}=L(T).
\]
$\widetilde{P}$ is usually called risk neutral probability in the financial
market. The process
\[
\widetilde{W}(t):=W(t)+\int_{0}^{t}\eta(s)\mathrm{d}s, \ 0\leq t\leq T
\]
is a Brownian motion under $\widetilde{P}$ by Girsanov's theorem.

Then we can rewrite the stock price processes (\ref{stock}) as
\[
dS_{i}(t)=S_{i}(t)\Big(\sum\limits_{j=1}^{d}\sigma_{ij}(t)d\widetilde{W}%
_{j}(t)\Big),\ i=1,...,d.
\]

Note that $\sigma(t)$ is assumed to be bounded, invertible and $\mathcal{G}%
_{t}$-adapted. So the filtration $\mathbb{G}$ coincides with the augmented
natural filtration of $\widetilde{W}$ by Theorem V.3.7 in \cite{Pro}.

Let $\hat\eta(t):=E[\eta(t)|\mathcal{G}_{t}]$ be a measurable version of the
conditional expectation of $\eta$ w.r.t. the filtration $\mathbb{G}$. Set
$\hat\mu$: $\hat\mu(t)=E[\mu(t)|\mathcal{G}_{t}]$. Then $\hat\mu
(t)=\sigma(t)\hat\eta(t)$, since $\sigma$ is $\mathbb{G}$-adapted.

We introduce the process
\begin{equation}
\widehat{W}(t):=\widetilde{W}(t)-\int_{0}^{t}\hat{\eta}(s)ds=W(t)+\int_{0}%
^{t}(\eta(s)-\hat{\eta}(s))ds,\ t\geq0.
\end{equation}

By Theorem 8.1.3 and Remark 8.1.1 in Kallianpur \cite{Ka}, $\{\widehat{W}%
(t),t\geq0\}$ is a $(\mathbb{G},P)$- Brownian motion which is the so-called
innovations process. Then, we could describe the dynamics of stock price
processes and the wealth process within a full observation model:
\[
dS_{i}(t)=S_{i}(t)\Big(\hat{\mu}_{i}(t)dt+\sum\limits_{j=1}^{d}\sigma
_{ij}(t)d\widehat{W}_{j}(t)\Big),\ i=1,...,d,
\]%
\[
dX(t)=\pi^{\prime}(t)\hat{\mu}(t)dt+\pi^{\prime}(t)\sigma(t)d\widehat{W}(t).
\]

Now all the coefficients in our model is observable. So we are in a full
observation model and our problem (\ref{optm}) can be reformulated as
\begin{equation}
\mathrm{Maximize}\ \ \ Y^{x_{0},\pi}(0),\label{optmf}\\
s.t.%
\begin{cases}
X(t)\geq0\ \ \forall t\in\lbrack0,T]\ \ a.s.,\\
\pi(\cdot)\in M_{\mathbb{G}}^{2},\\
(X(\cdot),\pi(\cdot))\ \ \ \mathrm{satisfies\ \ \ Eq.}(\ref{wealth1}),\\
(Y(\cdot),Z(\cdot))\ \ \ \mathrm{satisfies\ \ \ Eq.}(\ref{BSDE}).
\end{cases}
\end{equation}

\subsection{Backward formulation of the problem}

In this subsection, we first show BSDE (\ref{BSDE}) has a unique solution
under some mild conditions and then give an equivalent backward formulation of
problem (\ref{optmf}).

\begin{lemma}
Under Assumption \ref{assf1}, for $\forall\xi\in L_{\mathbb{G}}^{2}$, there
exists a unique solution $(Y,Z)$$\in S_{\mathbb{G}}^{2}\times M_{\mathbb{G}%
}^{2}$ to the BSDE (\ref{BSDE}).
\end{lemma}

Since $\mathbb{G}$ is strictly larger than the augmented filtration of the
$(P,\mathbb{G})$-Brownian motion $\widehat{W}$ in general, equation
(\ref{BSDE}) is not a standard BSDE. Fortunately, by Theorem 8.3.1 in
\cite{Ka}, every square integrable $\mathcal{G}_{t}$ -martingale $M(t)$ can be
represented as
\[
M(t)=M(0)+\int_{0}^{t}Z^{\prime}(s)d\widehat{W}(s),
\]
where $Z(\cdot)\in M_{\mathbb{G}}^{2}$. Thus, applying similar analysis as in
\cite{PP}, it is easy to prove this lemma.

Let $q(\cdot):=\sigma(\cdot)^{\prime}\pi(\cdot)$. Since $\sigma(\cdot)$ is
invertible, $q(\cdot)$ can be regarded as the control variable instead of
$\pi(\cdot)$. By the existence and uniqueness result of BSDE (\ref{BSDE}),
selecting $q(\cdot)$ is equivalent to selecting the terminal wealth $X(T)$. If
we take the terminal wealth as control variable, the wealth equation and
recursive utility process can be written as:
\begin{align}
\label{backsystem}%
\begin{cases}
-dX(t)=-q^{\prime-1}(t)\hat\mu(t)dt-q^{\prime}(t)d\widehat{W}(t),\\
X(T)=\xi,\\
-dY(t)=f(t,Y(t),Z(t))dt-Z^{\prime}(t)d\widehat{W}(t),\\
Y(T)=u(\xi),
\end{cases}
\end{align}
where the \textquotedblleft control" is the terminal wealth $\xi$ to be chosen
from the following set
\[
U:=\{\xi\big| \xi\in L^{2}_{\mathcal{G}_{T}}, \ \xi\geq0\}.
\]

From now on, we denote the solution of (\ref{backsystem}) by $(X^{\xi}%
(\cdot),q^{\xi}(\cdot),Y^{\xi}(\cdot),Z^{\xi}(\cdot))$. We also denote
$X^{\xi}(0)$ and $Y^{\xi}(0)$ by $X^{\xi}_{0}$ and $Y^{\xi}_{0}$ respectively.

As implied by the comparison theorem for BSDE (\ref{BSDE}), the nonnegative
terminal wealth,( $\xi=X(T)\geq0$) keeps the wealth process nonnegative all
the time. This gives rise to the following optimization problem:
\begin{equation}
\mathrm{Maximize}\ \ \ J(\xi):=Y_{0}^{\xi},\label{optmb}\\
s.t.%
\begin{cases}
\xi\in U,\\
X_{0}^{\xi}=x,\\
(X^{\xi}(\cdot),q^{\xi}(\cdot)),(Y^{\xi}(\cdot),Z^{\xi}(\cdot
))\ \ \ \mathrm{satisfies\ \ \ Eq.}(\ref{backsystem}).
\end{cases}
\end{equation}

\begin{definition}
A random variable $\xi\in U$ is called feasible for the initial wealth $x$ if
and only if $X^{\xi}(0)=x$. We will denote the set of all feasible $\xi$ for
the initial wealth $x$ by $\mathcal{A}(x)$.
\end{definition}


It is clear that original problems (\ref{optm}) and (\ref{optmf}) are
equivalent to the auxiliary one (\ref{optmb}). Hence, hereafter we focus
ourselves on solving (\ref{optmb}). Note that $\xi$ becomes the control
variable. The advantage of this approach is that the state constraint in
(\ref{optm}) becomes a control constraint in (\ref{optmb}), whereas it is well
known in control theory that a control constraint is much easier to deal with
than a state constraint. The cost of this approach is the original initial
condition $X^{\xi}(0)=x$ now becomes a constraint.

A feasible $\xi^{*}\in\mathcal{A}(x)$ is called optimal if it attains the
maximum of $J(\xi)$ over $\mathcal{A}(x)$. Once $\xi^{*}$ is determined, the
optimal portfolio is obtained by solving the first equation in
$(\ref{backsystem})$ with $X^{\xi^{*}}(T)=\xi^{*}$.

\section{Dual method for recursive utility maximization}

In this section, we impose the following concavity condition:

\begin{assumption}
\label{assfu} \text{The function} \ $(y,z)\mapsto f(\omega, t, y, z)$ \text{is
concave for all} \ $(\omega, t)\in\Omega\times[0,T]$.
\end{assumption}

We also need the following assumption on $u$:

\begin{assumption}
\label{inada} $u:(0,\infty)\rightarrow\mathbb{R}$ is strictly increasing,
strictly concave, continuously differentiable, and satisfies
\begin{equation}
u^{\prime}(0+):=\lim_{x\downarrow0}u^{\prime}(x)=\infty,\ \ \ \ u^{\prime
}(\infty):=\lim_{x\rightarrow\infty}u^{\prime}(x)=0.
\end{equation}

\end{assumption}

Under Assumption \ref{inada}, Assumption \ref{assu1} seems too restrictive and
it precludes some interesting examples. So in the following two sections, for
any given utility function $u$ satisfying Assumption \ref{inada}, we set
\[
U=\{\xi|\xi\in L_{\mathcal{G}_{T}}^{2},\ u(\xi)\in L_{\mathcal{G}_{T}}%
^{2}\ \text{and}\ \xi\geq0\}.
\]

In this section, we assume $\sigma\equiv I_{d}$, the d-dimensional identity
matrix. Let $F(t,\beta,\gamma)$ be the Fenchel-Legendre transform of $f$:
\begin{equation}
F(\omega,t,\beta,\gamma):=\sup_{(y,z)\in\mathbb{R}\times\mathbb{R}^{d}%
}\big[f(\omega,t,y,z)-y\beta-z^{\prime}\gamma\big],\ (\beta,\gamma
)\in\mathbb{R}\times\mathbb{R}^{d}.
\end{equation}
Let the effective domain of $F$ be $\mathcal{D}_{F}:=\{(\omega,t,\beta
,\gamma)\in\Omega\times\lbrack0,T]\times\mathbb{R}\times{\mathbb{R}}%
^{d}\big|F(\omega,t,\beta,\gamma)<+\infty\}$. As was shown in \cite{EPQ1}, the
$(\omega,t)$-section of $\mathcal{D}_{F}$, denoted by $\mathcal{D}%
_{F}^{(\omega,t)}$ is included in the bounded domain $B=[-C,C]^{d+1}%
\subset\mathbb{R}\times\mathbb{R}^{d}$, where $C$ is the Lipschitz constant of
$f$.

We have the duality relation by the concavity of $f$,
\begin{equation}
f(\omega, t,y,z)=\inf_{(\beta,\gamma)\in\mathcal{D}_{F}^{(\omega,t)}%
}\big[F(\omega, t,\beta,\gamma)+y\beta+z^{\prime}\gamma\big].
\end{equation}
For every $(\omega, t, y, z)$ the infimum is achieved in this relation by a
pair $(\beta, \gamma)$ which depends on $(\omega, t)$.

Set
\[
\mathcal{B}=\big\{(\beta,\gamma)\Big |(\beta,\gamma)\ \text{is}\ \mathbb{G}%
\text{-progressively measurable and B-valued}\ \text{and}\ E\int_{0}%
^{T}F(t,\beta_{t},\gamma_{t})^{2}dt<+\infty\big\}.
\]
Then $\mathcal{B}$ is a convex set. For any $(\beta,\gamma)\in\mathcal{B}$,
let
\[
f^{\beta,\gamma}(t,y,z)=F(t,\beta_{t},\gamma_{t})+y\beta_{t}+z^{\prime}%
\gamma_{t},
\]
and denote by $(Y^{\beta,\gamma},Z^{\beta,\gamma})$ the unique solution to the
linear BSDE (\ref{BSDE}) with $f^{\beta,\gamma}$.

By similar analysis as Proposition 3.4 in \cite{EPQ1}, we have the following
variational formulation of $X^{\xi}(t)$ and $Y^{\xi}(t)$.

\begin{lemma}
\label{rep} Under Assumption \ref{assf1} and \ref{assfu}, for any $\xi\in U$,
the solutions $(X^{\xi}(\cdot),q^{\xi}(\cdot)),(Y^{\xi}(\cdot),Z^{\xi}%
(\cdot))$ of Eq.(\ref{backsystem}) can be represented as
\begin{align*}
&  X^{\xi}(t)=\hat{L}^{-1}(t)E[\hat{L}(T)\xi|\mathcal{G}_{t}],\\
&  Y^{\xi}(t)=ess\inf_{{\beta,\gamma}\in\mathcal{B}}Y_{t}^{\beta,\gamma
},\ t\in\lbrack0,T],a.s.,
\end{align*}
where
\begin{align*}
&  \hat{L}(t):=e^{-\int_{0}^{t}\hat{\mu}^{\prime}(s)d\widehat{W}(s)-\frac
{1}{2}\int_{0}^{t}|\hat{\mu}(s)|^{2}ds},\\
&  Y_{t}^{\beta,\gamma}=E\big[\int_{t}^{T}\Gamma_{t,s}^{\beta,\gamma}%
F(s,\beta_{s},\gamma_{s})ds+\Gamma_{t,T}^{\beta,\gamma}u(\xi)|\mathcal{G}%
_{t}\big],\\
&  \Gamma_{t,s}^{\beta,\gamma}=e^{\int_{t}^{s}(\beta_{r}-\frac{1}{2}%
|\gamma_{r}|^{2})dr+\int_{t}^{s}\gamma_{r}^{\prime}d\widehat{W}(r)}.
\end{align*}
Especially, we have $Y^{\xi}(0)=\inf\limits_{(\beta,\gamma)\in\mathcal{B}%
}E\big[\int_{0}^{T}\Gamma_{0,s}^{\beta,\gamma}F(s,\beta_{s},\gamma
_{s})ds+\Gamma_{0,T}^{\beta,\gamma}u(\xi)\big]$.
\end{lemma}

\begin{remark}
By Theorem 3.1 in \cite{La}, we have $\hat{L}(t)=E[L(t)|\mathcal{G}%
_{t}],\ t\in\lbrack0,T],a.s..$
\end{remark}

By Lemma \ref{rep}, $\mathcal{A}(x)=\{\xi\in U\big|E[\hat{L}(T)\xi]= x\}$.
Thus, our problem is equivalent to the following problem:
\begin{align}
\mathrm{Maximize}\ \ \ J(\xi)=  &  \inf\limits_{\beta,\gamma\in\mathcal{B}%
}E\big[\int_{0}^{T}\Gamma_{0,s}^{\beta,\gamma}F(s,\beta_{s},\gamma
_{s})ds+\Gamma_{0,T}^{\beta,\gamma}u(\xi)\big]\nonumber\label{Kopt3}\\
&  s.t.\ \ \xi\in\mathcal{A}(x).
\end{align}

The maximum recursive utility that the investor can achieve is
\begin{equation}
\underline{V}(x):=\sup_{\xi\in\mathcal{A}(x)}\inf\limits_{\beta,\gamma
\in\mathcal{B}}E\big[\int_{0}^{T}\Gamma_{0,s}^{\beta,\gamma}F(s,\beta
_{s},\gamma_{s})ds+\Gamma_{0,T}^{\beta,\gamma}u(\xi)\big].
\end{equation}
It is dominated by its \textquotedblleft min-max" counterpart
\begin{equation}
\bar{V}(x):=\inf\limits_{(\beta,\gamma)\in\mathcal{B}}\sup_{\xi\in
\mathcal{A}(x)}E\big[\int_{0}^{T}\Gamma_{0,s}^{\beta,\gamma}F(s,\beta
_{s},\gamma_{s})ds+\Gamma_{0,T}^{\beta,\gamma}u(\xi)\big].
\end{equation}

If we can find $(\hat{\beta},\hat{\gamma},\hat{\xi})\in\mathcal{B}%
\times\mathcal{A}(x)$ such that
\begin{equation}
\underline{V}(x)=E\big[\int_{0}^{T}\Gamma_{0,s}^{\hat{\beta},\hat{\gamma}%
}F(s,\hat{\beta}_{s},\hat{\gamma}_{s})ds+\Gamma_{0,T}^{\hat{\beta},\hat
{\gamma}}u(\hat{\xi})\big]=\bar{V}(x), \label{game1}%
\end{equation}
then the optimal solution of problem (\ref{Kopt3}) is $\hat{\xi}$.

Let us introduce the monotone decreasing function $I(\cdot)$ as the inverse of
the marginal utility function $u^{\prime}(\cdot)$, and the convex dual
\begin{equation}
\tilde u(\zeta):=\max_{x> 0}[u(x)-\zeta x]=u(I(\zeta))-\zeta I(\zeta
),\ \zeta>0.
\end{equation}

Then, $\forall\xi\in\mathcal{A}(x),\ \forall(\beta,\gamma)\in\mathcal{B}%
,\ \ \forall\zeta>0,$
\begin{align}
\  &  \ \ \ \ E\big[\int_{0}^{T}\Gamma_{0,s}^{\beta,\gamma}F(s,\beta
_{s},\gamma_{s})ds+\Gamma_{0,T}^{\beta,\gamma}u(\xi
)\big]\nonumber\label{inqdua1}\\
&  \leq E\big[\int_{0}^{T}\Gamma_{0,s}^{\beta,\gamma}F(s,\beta_{s},\gamma
_{s})ds+\Gamma_{0,T}^{\beta,\gamma}\tilde{u}\big(\zeta\frac{\hat{L}(T)}%
{\Gamma_{0,T}^{\beta,\gamma}}\big)+\zeta\xi\hat{L}(T)\big]\nonumber\\
&  = E\big[\int_{0}^{T}\Gamma_{0,s}^{\beta,\gamma}F(s,\beta_{s},\gamma
_{s})ds+\Gamma_{0,T}^{\beta,\gamma}\tilde{u}\big(\zeta\frac{\hat{L}(T)}%
{\Gamma_{0,T}^{\beta,\gamma}}\big)\big]+\zeta x.
\end{align}

Furthermore, we have equality in the above formula for some $\hat{\xi}%
\in\mathcal{A}(x)$, $(\hat{\beta},\hat{\gamma})\in\mathcal{B}$, $\hat{\zeta
}>0$ if and only if the conditions
\begin{equation}
E[\hat{\xi}\hat{L}(T)]=x, \label{optcon11}%
\end{equation}%
\begin{equation}
\hat{\xi}=I\big(\hat{\zeta}\frac{\hat{L}(T)}{\Gamma_{0,T}^{\hat{\beta}%
,\hat{\gamma}}}\big),a.s. \label{optcon21}%
\end{equation}

are satisfied simultaneously. And in this case, we have
\begin{equation}
E\big[\int_{0}^{T}\Gamma_{0,s}^{\hat{\beta},\hat{\gamma}}F(s,\hat{\beta}%
_{s},\hat{\gamma}_{s})ds+\Gamma_{0,T}^{\hat{\beta},\hat{\gamma}}u(\hat{\xi
})\big]=E\big[\int_{0}^{T}\Gamma_{0,s}^{\hat{\beta},\hat{\gamma}}%
F(s,\hat{\beta}_{s},\hat{\gamma}_{s})ds+\Gamma_{0,T}^{\hat{\beta},\hat{\gamma
}}\tilde{u}\big(\hat{\zeta}\frac{\hat{L}(T)}{\Gamma_{0,T}^{\hat{\beta}%
,\hat{\gamma}}}\big)\big]+\hat{\zeta}x. \label{ff1}%
\end{equation}

\begin{lemma}
\label{suff1} Under Assumption \ref{assf1}, \ref{assfu} and \ref{inada},
suppose that there exists a quadruple $(\hat{\xi},\hat{\beta},\hat{\gamma
},\hat{\zeta})\in(\mathcal{A}(x)\times\mathcal{B}\times(0,\infty))$ which
satisfies (\ref{optcon11}), (\ref{optcon21}) and
\begin{equation}
E\big[\int_{0}^{T}\Gamma_{0,s}^{\hat{\beta},\hat{\gamma}}F(s,\hat{\beta}%
_{s},\hat{\gamma}_{s})ds+\Gamma_{0,T}^{\hat{\beta},\hat{\gamma}}u(\hat{\xi
})\big]\leq E\big[\int_{0}^{T}\Gamma_{0,s}^{\beta,\gamma}F(s,\beta_{s}%
,\gamma_{s})ds+\Gamma_{0,T}^{\beta,\gamma}u(\hat{\xi})\big],\ \forall
(\beta,\gamma)\in\mathcal{B}. \label{optcon31}%
\end{equation}
Then we have $\forall\xi\in\mathcal{A}(x),\ \forall(\beta,\gamma
)\in\mathcal{B}$,
\begin{align}
E\big[\int_{0}^{T}\Gamma_{0,s}^{\hat{\beta},\hat{\gamma}}F(s,\hat{\beta}%
_{s},\hat{\gamma}_{s})ds+\Gamma_{0,T}^{\hat{\beta},\hat{\gamma}}u(\xi)\big]
&  \leq E\big[\int_{0}^{T}\Gamma_{0,s}^{\hat{\beta},\hat{\gamma}}%
F(s,\hat{\beta}_{s},\hat{\gamma}_{s})ds+\Gamma_{0,T}^{\hat{\beta},\hat{\gamma
}}u(\hat{\xi})\big]\ \nonumber\label{saddlep1}\\
&  \leq E\big[\int_{0}^{T}\Gamma_{0,s}^{\beta,\gamma}F(s,\beta_{s},\gamma
_{s})ds+\Gamma_{0,T}^{\beta,\gamma}u(\hat{\xi})\big].
\end{align}
That is, $(\hat{\xi},\hat{\beta},\hat{\gamma})$ is a saddle point satisfying
(\ref{game1}).
\end{lemma}

\noindent\textbf{Proof:} We only prove the first relationship in
(\ref{saddlep1}). Let $(\beta,\gamma)=(\hat{\beta},\hat{\gamma})$ and
$\zeta=\hat{\zeta}$ in (\ref{inqdua1}). We get
\begin{align*}
&  \ \ \ E\big[\int_{0}^{T}\Gamma_{0,s}^{\hat{\beta},\hat{\gamma}}%
F(s,\hat{\beta}_{s},\hat{\gamma}_{s})ds+\Gamma_{0,T}^{\hat{\beta},\hat{\gamma
}}u(\xi)\big]\\
&  \leq E\big[\int_{0}^{T}\Gamma_{0,s}^{\hat{\beta},\hat{\gamma}}%
F(s,\hat{\beta}_{s},\hat{\gamma}_{s})ds+\Gamma_{0,T}^{\hat{\beta},\hat{\gamma
}}\tilde{u}\big(\hat{\zeta}\frac{\hat{L}(T)}{\Gamma_{0,T}^{\hat{\beta}%
,\hat{\gamma}}}\big)\big]+\hat{\zeta}x\\
&  =E\big[\int_{0}^{T}\Gamma_{0,s}^{\hat{\beta},\hat{\gamma}}F(s,\hat{\beta
}_{s},\hat{\gamma}_{s})ds+\Gamma_{0,T}^{\hat{\beta},\hat{\gamma}}u(\hat{\xi
})\big],\ \forall\xi\in\mathcal{A}(x),
\end{align*}
by $(\ref{ff1})$. This completes the proof. $\ \ \ \ \ \Box$

Let us introduce the value function
\begin{equation}
\tilde{V}(\zeta)\equiv\tilde{V}(\zeta;x):=\inf_{(\beta,\gamma)\in\mathcal{B}%
}E\big[\int_{0}^{T}\Gamma_{0,s}^{\beta,\gamma}F(s,\beta_{s},\gamma
_{s})ds+\Gamma_{0,T}^{\beta,\gamma}\tilde{u}\big(\zeta\frac{\hat{L}(T)}%
{\Gamma_{0,T}^{\beta,\gamma}}\big)\big],\ 0<\zeta<\infty. \label{dua2}%
\end{equation}
By (\ref{inqdua1}), we have
\begin{equation}
\bar{V}(x)\leq V_{\ast}(x), \label{dualeq1}%
\end{equation}
where
\begin{equation}
V_{\ast}(x):=\inf_{\zeta>0,(\beta,\gamma)\in\mathcal{B}}E\big[\int_{0}%
^{T}\Gamma_{0,s}^{\beta,\gamma}F(s,\beta_{s},\gamma_{s})ds+\Gamma_{0,T}%
^{\beta,\gamma}\tilde{u}\big(\zeta\frac{\hat{L}(T)}{\Gamma_{0,T}^{\beta
,\gamma}}\big)+\zeta x\big]=\inf_{\zeta>0}[\tilde{V}(\zeta)+\zeta x].
\label{dua11}%
\end{equation}

\begin{lemma}
\label{chaopt1} Under the assumptions of lemma \ref{suff1}, the followings
hold:\newline(\text{i}) $(\hat{\beta},\hat{\gamma})$ attains the infimum in
(\ref{dua2}) with $\zeta=\hat{\zeta}$.\newline(\text{ii}) The triple
$(\hat{\zeta},\hat{\beta},\hat{\gamma})$ attains the first infimum in
(\ref{dua11}).\newline(\text{iii}) The number $\hat{\zeta}\in(0,\infty)$
attains the second infimum in (\ref{dua11}).\newline(\text{iv}) There is no
\textquotedblleft duality gap" in (\ref{dualeq1}); that is,
\[
V_{\ast}(x)=\bar{V}(x)=\underline{V}(x)=E\big[\int_{0}^{T}\Gamma_{0,s}%
^{\hat{\beta},\hat{\gamma}}F(s,\hat{\beta}_{s},\hat{\gamma}_{s})ds+\Gamma
_{0,T}^{\hat{\beta},\hat{\gamma}}u(\hat{\xi})\big].
\]

\end{lemma}

\noindent\textbf{Proof:} (\text{i}) By $(\ref{ff1})$ and $(\ref{optcon31})$,
\begin{align*}
&  \ \ \ \ E\big[\int_{0}^{T}\Gamma_{0,s}^{\hat{\beta},\hat{\gamma}}%
F(s,\hat{\beta}_{s},\hat{\gamma}_{s})ds+\Gamma_{0,T}^{\hat{\beta},\hat{\gamma
}}\tilde{u}\big(\hat{\zeta}\frac{\hat{L}(T)}{\Gamma_{0,T}^{\hat{\beta}%
,\hat{\gamma}}}\big)\big]\\
&  =E\big[\int_{0}^{T}\Gamma_{0,s}^{\hat{\beta},\hat{\gamma}}F(s,\hat{\beta
}_{s},\hat{\gamma}_{s})ds+\Gamma_{0,T}^{\hat{\beta},\hat{\gamma}}u(\hat{\xi
})\big]-\hat{\zeta}x\\
&  \leq E\big[\int_{0}^{T}\Gamma_{0,s}^{\beta,\gamma}F(s,\beta_{s},\gamma
_{s})ds+\Gamma_{0,T}^{\beta,\gamma}u(\hat{\xi})\big]-\hat{\zeta}x\\
&  \leq E\big[\int_{0}^{T}\Gamma_{0,s}^{\beta,\gamma}F(s,\beta_{s},\gamma
_{s})ds+\Gamma_{0,T}^{\beta,\gamma}\tilde{u}\big(\hat{\zeta}\frac{\hat{L}%
(T)}{\Gamma_{0,T}^{\beta,\gamma}}\big)\big],\ \forall(\beta,\gamma
)\in\mathcal{B},
\end{align*}
where the last inequality is due to $(\ref{inqdua1})$.

(\text{ii}) By $(\ref{ff1})$ and $(\ref{optcon31})$, we have
\begin{align*}
&  \ \ \ \ E\big[\int_{0}^{T}\Gamma_{0,s}^{\hat{\beta},\hat{\gamma}}%
F(s,\hat{\beta}_{s},\hat{\gamma}_{s})ds+\Gamma_{0,T}^{\hat{\beta},\hat{\gamma
}}\tilde{u}\big(\hat{\zeta}\frac{\hat{L}(T)}{\Gamma_{0,T}^{\hat{\beta}%
,\hat{\gamma}}}\big)\big]+\hat{\zeta}x\\
&  =E\big[\int_{0}^{T}\Gamma_{0,s}^{\hat{\beta},\hat{\gamma}}F(s,\hat{\beta
}_{s},\hat{\gamma}_{s})ds+\Gamma_{0,T}^{\hat{\beta},\hat{\gamma}}u(\hat{\xi
})\big]\\
&  \leq E\big[\int_{0}^{T}\Gamma_{0,s}^{\beta,\gamma}F(s,\beta_{s},\gamma
_{s})ds+\Gamma_{0,T}^{\beta,\gamma}u(\hat{\xi})\big]\\
&  \leq E\big[\int_{0}^{T}\Gamma_{0,s}^{\beta,\gamma}F(s,\beta_{s},\gamma
_{s})ds+\Gamma_{0,T}^{\beta,\gamma}\tilde{u}\big(\zeta\frac{\hat{L}(T)}%
{\Gamma_{0,T}^{\beta,\gamma}}\big)\big]+\zeta x,\ \forall(\beta,\gamma
)\in\mathcal{B},\ \forall\zeta\in(0,\infty)
\end{align*}
where the last inequality is an application of $(\ref{inqdua1})$ to $\xi
=\hat{\xi}$.

(\text{iii}) By (\text{i}), $(\ref{ff1})$ and $(\ref{optcon31})$,
\begin{align*}
\tilde{V}(\hat{\zeta})+\hat{\zeta}x  &  =E\big[\int_{0}^{T}\Gamma_{0,s}%
^{\hat{\beta},\hat{\gamma}}F(s,\hat{\beta}_{s},\hat{\gamma}_{s})ds+\Gamma
_{0,T}^{\hat{\beta},\hat{\gamma}}\tilde{u}\big(\hat{\zeta}\frac{\hat{L}%
(T)}{\Gamma_{0,T}^{\hat{\beta},\hat{\gamma}}}\big)\big]+\hat{\zeta}x\\
&  =E\big[\int_{0}^{T}\Gamma_{0,s}^{\hat{\beta},\hat{\gamma}}F(s,\hat{\beta
}_{s},\hat{\gamma}_{s})ds+\Gamma_{0,T}^{\hat{\beta},\hat{\gamma}}u(\hat{\xi
})\big]\\
&  \leq E\big[\int_{0}^{T}\Gamma_{0,s}^{\beta,\gamma}F(s,\beta_{s},\gamma
_{s})ds+\Gamma_{0,T}^{\beta,\gamma}u(\hat{\xi})\big]\\
&  \leq E\big[\int_{0}^{T}\Gamma_{0,s}^{\beta,\gamma}F(s,\beta_{s},\gamma
_{s})ds+\Gamma_{0,T}^{\beta,\gamma}\tilde{u}\big(\zeta\frac{\hat{L}(T)}%
{\Gamma_{0,T}^{\beta,\gamma}}\big)\big]+\zeta x,\ \forall(\beta,\gamma
)\in\mathcal{B},\ \forall\zeta\in(0,\infty).
\end{align*}
So we get $\tilde{V}(\hat{\zeta})+\hat{\zeta}x\leq\inf\limits_{(\beta
,\gamma)\in\mathcal{B}}E\big[\int_{0}^{T}\Gamma_{0,s}^{\beta,\gamma}%
F(s,\beta_{s},\gamma_{s})ds+\Gamma_{0,T}^{\beta,\gamma}\tilde{u}%
\big(\zeta\frac{\hat{L}(T)}{\Gamma_{0,T}^{\beta,\gamma}}\big)\big]+\zeta
x=\tilde{V}(\zeta)+\zeta x,\ \forall\zeta\in(0,\infty)$.

(\text{iv}) By (\text{ii}) and $(\ref{ff1})$,
\begin{align*}
V_{\ast}(x)  &  =E\big[\int_{0}^{T}\Gamma_{0,s}^{\hat{\beta},\hat{\gamma}%
}F(s,\hat{\beta}_{s},\hat{\gamma}_{s})ds+\Gamma_{0,T}^{\hat{\beta},\hat
{\gamma}}\tilde{u}\big(\hat{\zeta}\frac{\hat{L}(T)}{\Gamma_{0,T}^{\hat{\beta
},\hat{\gamma}}}\big)\big]+\hat{\zeta}x\\
&  =E\big[\int_{0}^{T}\Gamma_{0,s}^{\hat{\beta},\hat{\gamma}}F(s,\hat{\beta
}_{s},\hat{\gamma}_{s})ds+\Gamma_{0,T}^{\hat{\beta},\hat{\gamma}}u(\hat{\xi
})\big]\\
&  =\bar{V}(x)=\underline{V}(x).
\end{align*}
This completes the proof. $\ \ \ \ \ \Box$

In the following, we prove the existence of the quadruple $(\hat{\xi}%
,\hat{\beta},\hat{\gamma},\hat{\zeta})$ which is postulated in Lemma
\ref{suff1}.

Notice that the function $x\mapsto x\tilde{u}(\frac{1}{x})$ is convex. By
similar analysis as in Appendix B of \cite{CC}, the following lemma holds.

\begin{lemma}
\label{hatnu11} Under Assumption \ref{assf1}, \ref{assfu} and \ref{inada}, for
any given $\zeta>0$, there exists a pair $(\hat{\beta},\hat{\gamma}%
)=(\hat{\beta}_{\zeta},\hat{\gamma}_{\zeta})\in\mathcal{B}$ which attains the
infimum in (\ref{dua2}).
\end{lemma}

\begin{lemma}
\label{hatzeta1} Under Assumption \ref{assf1}, \ref{assfu} and \ref{inada},
and suppose
\[
E\big[\int_{0}^{T}\Gamma_{0,s}^{\beta,\gamma}F(s,\beta_{s},\gamma
_{s})ds+\Gamma_{0,T}^{\beta,\gamma}\tilde{u}\big(\zeta\frac{\hat{L}(T)}%
{\Gamma_{0,T}^{\beta,\gamma}}\big)\big]<\infty,\ \forall\zeta>0,\ \forall
(\beta,\gamma)\in\mathcal{B}.
\]
Then for any given $x>0$, there exists a number $\hat{\zeta}=\hat{\zeta}%
_{x}\in(0,\infty)$ which attains $V_{\ast}(x)=\inf\limits_{\zeta>0}[\tilde
{V}(\zeta)+\zeta x].$
\end{lemma}

\noindent\textbf{Proof:} \textbf{Step 1:} By the convexity of $\tilde{u}$ and
Lemma \ref{hatnu11}, $\tilde{V}(\cdot)$ is convex. Fix $\zeta>0$, denote
$(\hat{\beta},\hat{\gamma})=(\hat{\beta}_{\zeta},\hat{\gamma}_{\zeta})$ as in
lemma \ref{hatnu11}. For any $\delta>0$, we have
\begin{align*}
\frac{\tilde{V}(\zeta+\delta)-\tilde{V}(\zeta)}{\delta}  &  \leq
\frac{E\big[\Gamma_{0,T}^{\hat{\beta},\hat{\gamma}}\tilde{u}\big((\zeta
+\delta)\frac{\hat{L}(T)}{\Gamma_{0,T}^{\hat{\beta},\hat{\gamma}}}%
\big)-\Gamma_{0,T}^{\hat{\beta},\hat{\gamma}}\tilde{u}\big(\zeta\frac{\hat
{L}(T)}{\Gamma_{0,T}^{\hat{\beta},\hat{\gamma}}}\big)\big]}{\delta}\\
&  \leq E\big[\hat{L}(T)\tilde{u}^{\prime}\big((\zeta+\delta)\frac{\hat{L}%
(T)}{\Gamma_{0,T}^{\hat{\beta},\hat{\gamma}}}\big)\big]=-E\big[\hat
{L}(T)I\big((\zeta+\delta)\frac{\hat{L}(T)}{\Gamma_{0,T}^{\hat{\beta}%
,\hat{\gamma}}}\big)\big].
\end{align*}
Then, by Levi's lemma,
\begin{equation}
\lim_{\delta\rightarrow0+}\frac{\tilde{V}(\zeta+\delta)-\tilde{V}(\zeta
)}{\delta}\leq-E\big[\hat{L}(T)I\big(\zeta\frac{\hat{L}(T)}{\Gamma_{0,T}%
^{\hat{\beta},\hat{\gamma}}}\big)\big]
\end{equation}
and
\begin{equation}
\lim_{\delta\rightarrow0+}\frac{\tilde{V}(\zeta)-\tilde{V}(\zeta-\delta
)}{\delta}\geq-E\big[\hat{L}(T)I\big(\zeta\frac{\hat{L}(T)}{\Gamma_{0,T}%
^{\hat{\beta},\hat{\gamma}}}\big)\big].
\end{equation}
Since $\tilde{V}(\cdot)$ is convex, we obtain that $\tilde{V}(\cdot)$ is
differentiable on $(0,\infty)$ and $\tilde{V}^{\prime}(\zeta)=-E\big[\hat
{L}(T)I\big(\zeta\frac{\hat{L}(T)}{\Gamma_{0,T}^{\hat{\beta},\hat{\gamma}}%
}\big)\big]$.

\textbf{Step 2:} Because $\mu(\cdot)$ is bounded, we have that for any
$\zeta\in(0, \infty)$, $\frac{\hat{L}(T)}{\Gamma_{0,T}^{\hat{\beta}%
,\hat{\gamma}}}<+\infty,a.s.$. Then,
\[
\tilde{V}^{\prime}(+\infty):=\lim_{\zeta\rightarrow+\infty}\tilde{V}^{\prime
}(\zeta)=-\lim_{\zeta\rightarrow\infty}E\big[\hat{L}(T)I\big(\zeta\frac
{\hat{L}(T)}{\Gamma_{0,T}^{\hat{\beta},\hat{\gamma}}}\big)\big]=0,
\]
\[
\tilde{V}^{\prime}(0):=\lim_{\zeta\rightarrow0+}\tilde{V}^{\prime}%
(\zeta)=-\lim_{\zeta\rightarrow0+}E\big[\hat{L}(T)I\big(\zeta\frac{\hat{L}%
(T)}{\Gamma_{0,T}^{\hat{\beta},\hat{\gamma}}}\big)\big]=-\infty.
\]

Thus, there exists a number $\hat{\zeta}$ which attains $V_{\ast}(x)$ and
$\tilde{V}^{\prime}(\hat{\zeta})=-x\in(-\infty,0)$. This completes the proof.
$\ \ \ \ \ \Box$

\begin{lemma}
\label{hatnuzeta1} Under Assumption \ref{assf1}, \ref{assfu} and \ref{inada},
$V_{*}(x)=E\big[\int_{0}^{T}\Gamma_{0,s}^{\hat\beta,\hat\gamma}F(s,\hat
\beta_{s},\hat\gamma_{s})ds+\Gamma_{0,T}^{\hat\beta,\hat\gamma}u\big(\hat
\zeta\frac{\hat L(T)}{\Gamma_{0,T}^{\hat\beta,\hat\gamma}}\big)\big]+\hat\zeta
x$ with $\hat\zeta=\hat\zeta_{x}$ as in lemma \ref{hatzeta1} and $(\hat\beta,
\hat\gamma)=(\hat\beta_{\hat\zeta}, \hat\gamma_{\hat\zeta})$ as in lemma
\ref{hatnu11}.
\end{lemma}

\noindent\textbf{Proof:} We have
\begin{align*}
&  \ \ \ \ E\big[\int_{0}^{T}\Gamma_{0,s}^{\hat{\beta},\hat{\gamma}}%
F(s,\hat{\beta}_{s},\hat{\gamma}_{s})ds+\Gamma_{0,T}^{\hat{\beta},\hat{\gamma
}}\tilde{u}\big(\hat{\zeta}\frac{\hat{L}(T)}{\Gamma_{0,T}^{\hat{\beta}%
,\hat{\gamma}}}\big)\big]+\hat{\zeta}x\\
&  =\tilde{V}(\hat{\zeta})+\hat{\zeta}x\\
&  \leq\tilde{V}({\zeta})+\zeta x\\
&  \leq E\big[\int_{0}^{T}\Gamma_{0,s}^{\beta,\gamma}F(s,\beta_{s},\gamma
_{s})ds+\Gamma_{0,T}^{\beta,\gamma}\tilde{u}\big(\zeta\frac{\hat{L}(T)}%
{\Gamma_{0,T}^{\beta,\gamma}}\big)+\zeta x\big],\ \forall(\beta,\gamma
)\in\mathcal{B},\ \forall\zeta\in(0,\infty),\ \forall x>0.
\end{align*}
This completes the proof. $\ \ \ \ \ \Box$

Our main result is the following theorem.

\begin{theorem}
\label{main} Under Assumption \ref{assf1}, \ref{assfu} and \ref{inada}, let
$(\hat{\zeta},\hat{\beta},\hat{\gamma})$ as in lemma \ref{hatnuzeta1} and
define $\hat{\xi}=I\big(\hat{\zeta}\frac{\hat{L}(T)}{\Gamma_{0,T}^{\hat{\beta
},\hat{\gamma}}}\big)\ a.s.$. If $\hat{\xi}\in U$, then $(\hat{\zeta}%
,\hat{\beta},\hat{\gamma},\hat{\xi})$ satisfies all the conditions in lemma
\ref{suff1}, that is (\ref{optcon11}), (\ref{optcon21}) and (\ref{optcon31}).
\end{theorem}

\noindent\textbf{Proof: }Notice that
\[
\tilde{V}(\hat{\zeta})=\inf\limits_{(\beta,\gamma)\in\mathcal{B}}%
E\big[\int_{0}^{T}\Gamma_{0,s}^{\beta,\gamma}F(s,\beta_{s},\gamma
_{s})ds+\Gamma_{0,T}^{\beta,\gamma}\tilde{u}\big(\hat{\zeta}\frac{\hat{L}%
(T)}{\Gamma_{0,T}^{\beta,\gamma}}\big)\big]=E\big[\int_{0}^{T}\Gamma
_{0,s}^{\hat{\beta},\hat{\gamma}}F(s,\hat{\beta}_{s},\hat{\gamma}%
_{s})ds+\Gamma_{0,T}^{\hat{\beta},\hat{\gamma}}\tilde{u}\big(\hat{\zeta}%
\frac{\hat{L}(T)}{\Gamma_{0,T}^{\hat{\beta},\hat{\gamma}}}\big)\big].
\]
Applying the maximum principle in Peng \cite{Pe}, we obtain a necessary
condition for $(\hat{\beta},\hat{\gamma})$:
\begin{equation}
F(t,\beta_{t},\gamma_{t})+p_{t}\beta_{t}+q_{t}\gamma_{t}\geq F(t,\hat{\beta
}_{t},\hat{\gamma}_{t})+p_{t}\hat{\beta}_{t}+q_{t}\hat{\gamma}_{t}%
,\ \forall(\beta,\gamma)\in\mathcal{B}, \label{qmp1}%
\end{equation}
where $(p_{t},q_{t})$ is the solution of the adjoint equation%

\begin{equation}%
\begin{cases}
-dp_{t}=\big(F(t,\hat\beta_{t}, \hat\gamma_{t})+p_{t}\hat\beta_{t}+q^{\prime
}_{t}\hat\gamma_{t}\big)dt-q^{\prime}_{t}d\widehat{W}(t),\\
p_{T}=u\big(I(\hat\zeta\frac{\hat L(T)}{\Gamma_{0,T}^{\hat\beta, \hat\gamma}%
})\big).
\end{cases}
\end{equation}

$\forall(\beta,\gamma)\in\mathcal{B}$, let $(y_{t},z_{t})$ and $(\tilde{y}%
_{t},\tilde{z}_{t})$ be the unique solutions of the following two linear
BSDEs, respectively,%

\begin{equation}
y_{t}=u(\hat{\xi})+\int_{t}^{T}\big(y_{s}\hat{\beta}_{s}+z_{s}^{\prime}%
\hat{\gamma}_{s}+F(s,\hat{\beta}_{s},\hat{\gamma}_{s})\big)ds-\int_{t}%
^{T}z_{s}^{\prime}d\widehat{W}(s), \label{optu}%
\end{equation}%
\begin{equation}
\tilde{y}_{t}=u(\hat{\xi})+\int_{t}^{T}\big(\tilde{y}_{s}\beta_{s}+\tilde
{z}_{s}^{\prime}\gamma_{s}+F(s,\beta_{s},\gamma_{s})\big)ds-\int_{t}^{T}%
\tilde{z}_{s}^{\prime}d\widehat{W}(s).
\end{equation}

By (\ref{qmp1}) and the comparison theorem of BSDE, we have $y_{t}\leq\tilde
y_{t}, \ t\in[0,T], a.s.$, especially $y_{0}\leq\tilde y_{0}.$

Solving the above linear BSDEs gives
\[
y_{0}=E\big[\int_{0}^{T}\Gamma_{0,s}^{\hat{\beta},\hat{\gamma}}F(s,\hat{\beta
}_{s},\hat{\gamma}_{s})ds+\Gamma_{0,T}^{\hat{\beta},\hat{\gamma}}u(\hat{\xi
})\big]
\]
and
\[
\tilde{y}_{0}=E\big[\int_{0}^{T}\Gamma_{0,s}^{\beta,\gamma}F(s,\beta
_{s},\gamma_{s})ds+\Gamma_{0,T}^{\beta,\gamma}u(\hat{\xi})\big].
\]

So
\begin{equation}
E\big[\int_{0}^{T}\Gamma_{0,s}^{\hat{\beta},\hat{\gamma}}F(s,\hat{\beta}%
_{s},\hat{\gamma}_{s})ds+\Gamma_{0,T}^{\hat{\beta},\hat{\gamma}}u(\hat{\xi
})\big]\leq E\big[\int_{0}^{T}\Gamma_{0,s}^{\beta,\gamma}F(s,\beta_{s}%
,\gamma_{s})ds+\Gamma_{0,T}^{\beta,\gamma}u(\hat{\xi})\big],\ \forall
(\beta,\gamma)\in\mathcal{B},\nonumber
\end{equation}
which exactly is Eq.(\ref{optcon31}).

By Lemma \ref{hatzeta1}, $\tilde{V}^{\prime}(\hat{\zeta})=-x$. By Lemma
\ref{hatnu11},
\begin{equation}
\tilde{V}(\hat{\zeta})=E\big[\int_{0}^{T}\Gamma_{0,s}^{\hat{\beta},\hat
{\gamma}}F(s,\hat{\beta}_{s},\hat{\gamma}_{s})ds+\Gamma_{0,T}^{\hat{\beta
},\hat{\gamma}}\tilde{u}\big(\hat{\zeta}\frac{\hat{L}(T)}{\Gamma_{0,T}%
^{\hat{\beta},\hat{\gamma}}}\big)\big]. \label{hatzetap}%
\end{equation}
Differentiating both sides of (\ref{hatzetap}) as functions of $\hat{\zeta}$,
we get
\begin{equation}
E[I\big(\hat{\zeta}\frac{\hat{L}(T)}{\Gamma_{0,T}^{\hat{\beta},\hat{\gamma}}%
}\big)\hat{L}(T)]=x.
\end{equation}
This completes the proof. $\ \ \ \ \ \Box$

\begin{remark}
It is worth to pointing out that the adjoint process $p_{t}$ in the proof of
the above theorem coincides with the optimal utility process $y_{t}$ in
Eq.(\ref{optu}).
\end{remark}

\section{K-ignorance}

In this section, we study a special case which is called K-ignorance by Chen
and Epstein \cite{CE}. In this case, the generator $f$ is specified as
\[
f(t,y,z)=-K|z|,\ K\geq0.
\]
Chen and Epstein interpreted the term $K|z|$ as modeling ambiguity aversion
rather than risk aversion. $f(z)=-K|z|$ is not differentiable. But it is
concave and $f(z)=\inf\limits_{|\gamma|\leq K}(\gamma z)$. Then, our results
in the above section are still applicable.

In this section, we assume $d=1$, $\sigma\equiv1$. The wealth equation and
recursive utility become
\begin{equation}%
\begin{cases}
-dX(t)=-q^{\prime}(t)\hat{\mu}(t)dt-q^{\prime}(t)d\widehat{W}(t),\\
X(T)=\xi,\\
-dY(t)=-K|Z(t)|dt-Z^{\prime}(t)d\widehat{W}(t),\\
Y(T)=u(\xi).
\end{cases}
\label{Ksystem}%
\end{equation}
Our problem is formulated as:
\begin{equation}
\mathrm{Maximize}\ \ \ J(\xi):=Y_{0}^{\xi},\label{Kopt}\\
s.t.%
\begin{cases}
\xi\in U,\\
X(0)=x,\\
(X(\cdot),q(\cdot)),(Y(\cdot),Z(\cdot))\ \ \ \mathrm{satisfies\ \ \ Eq.}%
(\ref{Ksystem}).
\end{cases}
\end{equation}

Now Lemma \ref{rep} can be simplified to the following lemma.

\begin{lemma}
For $\xi\in U$, the solutions $(X(\cdot),q(\cdot))$ and $(Y(\cdot),Z(\cdot))$
of Eq.(\ref{Ksystem}) can be represented as
\begin{align*}
&  X(t)=\hat{L}^{-1}(t)E[\hat{L}(T)\xi|\mathcal{G}_{t}],\\
&  Y(t)=ess\inf_{\gamma\in\mathcal{B}}(\Gamma_{0,t}^{0,\gamma})^{-1}%
(t)E[\Gamma_{0,T}^{0,\gamma}u(\xi)|\mathcal{G}_{t}],
\end{align*}
where
\begin{align*}
&  \hat{L}(t)=e^{-\int_{0}^{t}\hat{\mu}(s)\mathrm{d}\widehat{W}(s)-\frac{1}%
{2}\int_{0}^{t}|\hat{\mu}(s)|^{2}ds},\\
&  \Gamma_{0,t}^{0,\gamma}=e^{\int_{0}^{t}\gamma_{s}\mathrm{d}\widehat{W}%
(s)-\frac{1}{2}\int_{0}^{t}|\gamma_{s}|^{2}ds},\\
&  \mathcal{B}=\{\gamma=\{\gamma_{t}\}_{t\geq0}|\gamma_{t}\ is\ \mathbb{G}%
\text{-progressively measurable},\ |\gamma_{t}|\leq K,\ t\in\lbrack
0,T],a.s.\}.
\end{align*}

\end{lemma}

For any $\gamma\in\mathcal{B}$, $\Gamma_{0,t}^{0,\gamma}$ is $(\mathbb{G}%
,P)$-martingale. Then, a new probability measure $P_{\gamma}$ is defined on
$\mathcal{G}_{T}$ by
\[
\frac{dP_{\gamma}}{dP}=\Gamma_{0,T}^{0,\gamma}%
\]
and $\widehat{W}_{\gamma}(t)=\widehat{W}(t)-\int_{0}^{t}\gamma_{s}ds$ is a
Brownian motion under $P_{\gamma}$. Thus, $Y(0)=\inf\limits_{\gamma
\in\mathcal{B}}E_{\gamma}[u(\xi)]$ where $E_{\gamma}[\cdot]$ is the
expectation operator with respect to $P_{\gamma}$.

Our problem (\ref{Kopt}) is equivalent to the following problem:%
\begin{align}
\mathrm{Maximize}\ J(\xi)  &  =\inf_{\gamma\in\mathcal{B}}E_{\gamma}%
u(\xi)\nonumber\label{Kopt2}\\
s.t.\ \xi &  \in\mathcal{A}(x).
\end{align}
The auxiliary dual problem in (\ref{dua2}) becomes
\begin{equation}
\tilde{V}(\zeta)\equiv\tilde{V}(\zeta;x):=\inf_{\gamma\in\mathcal{B}}%
E_{\gamma}\tilde{u}(\zeta Z_{\gamma}(T)),\ 0<\zeta<\infty, \label{dua}%
\end{equation}
where $Z_{\gamma}(t):=\frac{\hat{L}(t)}{\Gamma_{0,t}^{0,\gamma}},\ t\in
\lbrack0,T],$ $a.s.$ and
\begin{equation}
V_{\ast}(x):=\inf_{\zeta>0,\gamma\in\mathcal{B}}[E_{\gamma}\tilde{u}(\zeta
Z_{\gamma}(T))+\zeta x]=\inf_{\zeta>0}[\tilde{V}(\zeta)+\zeta x]. \label{dua1}%
\end{equation}

Applying the procedure in the previous section, we can find the saddle point.
So we list the results without proof except Lemma \ref{hatnu} in which a new
proof is given.

\begin{lemma}
\label{hatnu} Under Assumption \ref{inada}, for any given $\zeta>0$, there
exists a unique $\hat{\gamma}=\hat{\gamma}_{\zeta}\in\mathcal{B}$ which
attains the infimum in (\ref{dua}).
\end{lemma}

\noindent\textbf{Proof:} Set $\mathcal{B}^{\prime}=\{\Gamma_{0,T}^{0,\gamma
}\mid\gamma\in\mathcal{B}\}$, $\mathcal{M}=\{M_{\gamma}(T):=\frac{\Gamma
_{0,T}^{0,\gamma}}{\hat{L}(T)}\mid\gamma\in\mathcal{B}\}$ and $g(x)=x\tilde
{u}(\frac{\zeta}{x})$ for $x>0$. Then problem $(\ref{dua})$ becomes
\begin{equation}
\tilde{V}({\zeta})=\inf\limits_{M_{\gamma}(T)\in\mathcal{M}}\tilde
{E}[M_{\gamma}(T)\tilde{u}(\zeta\frac{1}{M_{\gamma}(T)})]=\inf
\limits_{M_{\gamma}(T)\in\mathcal{M}}\tilde{E}[g(M_{\gamma}(T))]
\label{hatnu1}%
\end{equation}
where $\tilde{E}[\cdot]$ is the expectation operator w.r.t. the risk neutral
measure $\tilde{P}$. By Theorem 2.1 in \cite{CE}, we know $\mathcal{B}%
^{\prime}$ is norm closed in $L^{1}(\Omega)$. So $\mathcal{B}$ is closed under
a.s. convergence because $B$ is uniformly integrable. As a consequence,
$\mathcal{M}$ is closed under a.s. convergence.

Consider a minimizing sequence $\{M_{\gamma^{n}}(T)\}_{n\geq1}$ for
($\ref{hatnu1}$), that is
\[
\lim\limits_{n\rightarrow\infty}\tilde{E}[g(M_{\gamma^{n}}(T))]=\tilde
{V}(\zeta).
\]
By Komlos' theorem, there exists a sequence $\bar{M}_{\gamma^{n}}(T)\in$
conv$(M_{\gamma^{n}}(T),M_{\gamma^{n+1}}(T),...)$, i.e. $\bar{M}_{\gamma^{n}%
}(T)=\sum_{k=n}^{T_{n}}\lambda_{k}M_{\gamma^{k}}(T)$, $\lambda_{k}\in
\lbrack0,1]$ and $\sum_{k=n}^{T_{n}}\lambda_{k}=1$, such that the sequence
$\{M_{\gamma^{n}}(T)\}_{n\geq1}$ converges a.s. to a random variable $M$. By
the a.s. closedness of $\mathcal{M}$, we have $M\in\mathcal{M}$, that is
$\exists\hat{\gamma}\in\mathcal{B}$, s.t. $M=M_{\hat{\gamma}}(T)$. Note that
$g$ is a strictly convex continuous function, we have
\begin{align*}
\tilde{E}[g(M)]  &  =\tilde{E}[\lim_{n\rightarrow\infty}g(\bar{M}_{\gamma^{n}%
}(T))]\leq\liminf_{n\rightarrow\infty}\tilde{E}[g(\bar{M}_{\gamma^{n}}(T))]\\
&  \leq\liminf_{n\rightarrow\infty}\lambda_{k}\sum_{k=n}^{T_{n}}\tilde
{E}[g(M_{\gamma^{k}}(T))]=\liminf_{n\rightarrow\infty}\tilde{E}[g(M_{\gamma
^{n}}(T))]=\tilde{V}(\zeta).
\end{align*}
The uniqueness follows from the strictly convexity of $g$. This completes the
proof. $\ \ \ \ \ \Box$

\begin{lemma}
\label{hatzeta} Under Assumption \ref{inada}, if $\tilde{E}[I(\zeta Z_{\gamma
}(T))]<\infty,\ \forall\zeta>0,\ \forall\gamma\in\mathcal{B}$, then for any
given $x>0$, there exists a number $\hat{\zeta}=\hat{\zeta}_{x}\in(0,\infty)$
which attains the infimum of $V_{\ast}(x)=\inf\limits_{\zeta>0}[\tilde
{V}(\zeta)+\zeta x]$.
\end{lemma}

\begin{lemma}
\label{hatnuzeta} Under Assumption \ref{inada}, $V_{*}(x)=E_{\hat\gamma}
\tilde u(\hat\zeta Z_{\hat\gamma}(T))+\hat\zeta x$ with $\hat\zeta=\hat
\zeta_{x}$ as in lemma \ref{hatzeta} and $\hat\gamma=\hat\gamma_{\hat\zeta}$
as in lemma \ref{hatnu}.
\end{lemma}

\begin{theorem}
\label{mainth} Under Assumption \ref{inada}, let $(\hat{\zeta},\hat{\gamma})$
is the same as in lemma \ref{hatnuzeta}, then the optimal terminal wealth of
problem (\ref{Kopt2}) is
\[
\hat{\xi}=I(\hat{\zeta}Z_{\hat{\gamma}}(T)),a.s.
\]
if $\hat{\xi}$ belongs to $U$.
\end{theorem}

In the following, we give some examples to illustrate our above analysis.

\begin{example}
(Constant absolute risk aversion). Suppose that $u(x)=1-e^{-\alpha x}%
,\ x\in\mathbb{R},\ \alpha>0$, and the wealth of the investor may be negative.
This utility function $u$ does not satisfies Assumption \ref{inada}. But it
satisfies the following assumption:

\begin{assumption}
\label{inada1} $u$ is strictly increasing, strictly concave, continuously
differentiable, and
\begin{equation}
u^{\prime}(-\infty):=\lim_{x\downarrow-\infty}u^{\prime}(x)=\infty
,\ \ \ \ u^{\prime}(\infty):=\lim_{x\rightarrow\infty}u^{\prime}(x)=0.
\end{equation}

\end{assumption}

Note that under Assumption \ref{inada1}, the results in this section still hold.

For this example, $I(\zeta)=-\frac{1}{\alpha}\ln\frac{\zeta}{\alpha}%
,\ \zeta>0$, and $\tilde{u}(\zeta)=1-\frac{\zeta}{\alpha}+\frac{\zeta}{\alpha
}\ln\frac{\zeta}{\alpha},\ \zeta>0$. Then the value function of the auxiliary
dual problem (\ref{dua}) is
\begin{align*}
E_{\gamma}\tilde{u}(\zeta Z_{\gamma}(T))  &  =1-\frac{\zeta}{\alpha}%
+\frac{\zeta}{\alpha}\ln\frac{\zeta}{\alpha}+\frac{\zeta}{\alpha}\tilde{E}[\ln
Z_{\gamma}(T)]\\
&  =1-\frac{\zeta}{\alpha}+\frac{\zeta}{\alpha}\ln\frac{\zeta}{\alpha}%
+\frac{\zeta}{2\alpha}\tilde{E}\int_{0}^{T}(\hat{\mu}(t)+\gamma_{t}%
)^{2}dt,\ \zeta>0.
\end{align*}
Apparently, $\hat{\gamma}_{t}$ (the optimal $\gamma_{t}$) which attains the
infimum of Problem (\ref{dua}) is independent of $\zeta$. It is easy to see
that
\[
\hat{\gamma}_{t}=(-K)\vee(-\hat{\mu}(t))\wedge K,\ t\in\lbrack0,T],a.s..
\]
The optimal value of Problem (\ref{dua}) is
\[
\tilde{V}(\zeta)=1-\frac{\zeta}{\alpha}+\frac{\zeta}{\alpha}\ln\frac{\zeta
}{\alpha}+\frac{\zeta}{2\alpha}\tilde{E}\int_{0}^{T}(\hat{\mu}(t)+\hat{\gamma
}_{t})^{2}dt,
\]
and the Lagrange multiplier in Lemma \ref{hatzeta} is
\[
\hat{\zeta}\equiv\hat{\zeta}_{x}=\alpha e^{-\frac{1}{2}\tilde{E}\int_{0}%
^{T}(\hat{\mu}(t)+\hat{\gamma}_{t})^{2}dt-\alpha x}=\arg\min_{\zeta>0}%
[\tilde{V}(\zeta)+\zeta x].
\]

Thus, the optimal terminal wealth in Theorem \ref{mainth} is
\[
\hat{\xi}=-\frac{1}{\alpha}\ln\frac{\hat{\zeta}Z_{\hat{\gamma}}(T)}{\alpha}.
\]

Moreover, it is easy to check that $(Y(t),Z(t)):=\big(1-\frac{\hat{\zeta}%
}{\alpha}Z_{\hat{\gamma}}(t),\frac{\hat{\zeta}}{\alpha}(\hat{\mu}%
(t)+\hat{\gamma}_{t})Z_{\hat{\gamma}}(t)\big),\ t\in\lbrack0,T]$ uniquely
solves the utility equation in Eq.(\ref{Ksystem}) when $\xi=\hat{\xi}$.
\end{example}

\begin{example}
(Logarithmic utility function) Suppose $u(x)=\ln x,\ x>0.$ In this case,
\[
I(\zeta)=\frac{1}{\zeta},\ \ \text{and}\ \ \tilde{u}(\zeta)=-\ln
\zeta-1,\ \zeta>0.
\]
Then the value function of the auxiliary dual problem (\ref{dua}) is
\begin{align*}
E_{\gamma}\tilde{u}(\zeta Z_{\gamma}(T))  &  =E_{\gamma}[-\ln(\zeta Z_{\gamma
}(T))-1]\\
&  =E_{\gamma}[-\ln Z_{\gamma}(T)]-\ln\zeta-1\\
&  =\frac{1}{2}E_{\gamma}\int_{0}^{T}(\hat{\mu}(t)+\gamma_{t})^{2}dt-\ln
\zeta-1,\ \zeta>0.
\end{align*}
So the optimal $\hat{\gamma}_{t}$ is independent of $\zeta$. Consider the
following BSDE
\[
y_{\gamma}(t)=E_{\gamma}[\int_{t}^{T}(\hat{\mu}(s)+\gamma_{s})^{2}%
ds\big|\mathcal{G}_{t}]=\int_{t}^{T}[(\hat{\mu}(s)+\gamma_{s})^{2}+\gamma
_{s}z_{\gamma}(s)]ds-\int_{t}^{T}z_{\gamma}(s)d\widehat{W}(s).
\]

Set
\begin{align*}
f(t,z_{t})  &  =\inf\limits_{\gamma\in\mathcal{B}}[(\hat{\mu}(t)+\gamma
_{t})^{2}+\gamma_{t}z_{t}]\\
&  =%
\begin{cases}
K^{2}-2K\hat{\mu}(t)-Kz_{t}+\hat{\mu}(t)^{2},\ \ \ \text{if}-2\hat{\mu
}(t)+2K<z_{t};\\
-\frac{1}{4}z_{t}^{2}-\hat{\mu}(t)z_{t}%
,\ \ \ \ \ \ \ \ \ \ \ \ \ \ \ \ \ \ \ \ \text{if}-2\hat{\mu}(t)-2K\leq
z_{t}\leq-2\hat{\mu}(t)+2K;\\
K^{2}+2K\hat{\mu}(t)+Kz_{t}+\hat{\mu}(t)^{2},\ \ \ \text{if}\ z_{t}<-2\hat
{\mu}(t)-2K,\ t\in\lbrack0,T],a.s..
\end{cases}
\end{align*}
It is easy to show that $f(t, z_t)$ is uniformly Lipschitz, so the following BSDE has a
unique solution which we still denoted by $(y_{t},z_{t})$.
\begin{equation}
y_{t}=\int_{t}^{T}f(s,z_{s})ds-\int_{t}^{T}z_{s}d\widehat{W}(s).
\end{equation}
Then the infimum in problem (\ref{dua}) is attained at
\begin{align*}
\hat{\gamma}_{t}  &  =\arg\inf\limits_{\gamma\in\mathcal{B}}[(\hat{\mu
}(t)+\gamma_{t})^{2}+\gamma_{t}z_{t}]\\
&  =-KI_{\{-2\hat{\mu}(t)+2K<z_{t}\}}+(-\hat{\mu}(t)-\frac{z_{t}}%
{2})I_{\{-2\hat{\mu}(t)-2K\leq z_{t}\leq-2\hat{\mu}(t)+2K\}}+KI_{\{z_{t}%
<-2\hat{\mu}(t)-2K\}},\ t\in\lbrack0,T],a.s..
\end{align*}

The Lagrange multiplier in Lemma \ref{hatzeta} is
\[
\hat{\zeta}\equiv\hat{\zeta}_{x}=\frac{1}{x}=\arg\min_{\zeta>0}[\tilde
{V}(\zeta)+\zeta x].
\]

The optimal terminal wealth in Theorem \ref{mainth} is
\[
\hat{\xi}=\frac{x}{Z_{\hat{\gamma}}(T)}.
\]

\end{example}

\begin{example}
Suppose that the appreciation rate $\mu(t)$ is a bounded deterministic
function of $t$. In this case, $\mathcal{G}_{t}=\mathcal{F}_{t},\ t\geq0$, and
we claim that
\begin{equation}
\hat{\gamma}_{t}=(-K)\vee(-\mu(t))\wedge K,\ t\in\lbrack0,T]. \label{hatnu2}%
\end{equation}
\noindent\textbf{Proof:} We show that $\hat{\gamma}$ defined above attains the
infimum of (\ref{dua}). Denote
\[
v(t,x)\equiv v(t,x;\zeta):=\tilde{E}[g(x\frac{M_{\hat{\gamma}}(T)}%
{M_{\hat{\gamma}}(t)})].
\]
Then $v(t,x)$ is the solution of the partial differential equation
$\frac{\partial v}{\partial t}+\frac{1}{2}\frac{\partial^{2}v}{\partial x^{2}%
}x^2(\mu_t+\hat\gamma_t)^2=0.$

$\forall\gamma\in\mathcal{B}$, applying It\^{o}'s formula to $v(t,M_{\gamma
}(t))$, we have
\begin{align*}
dv(t,M_{\gamma}(t)) &  =[\frac{\partial v}{\partial t}+\frac{1}{2}%
\frac{\partial^{2}v}{\partial x^{2}}(M_{\gamma}(t))^{2}(\mu(t)+\gamma_{t}%
)^{2}]dt+\frac{\partial v}{\partial x}M_{\gamma}(t)(\mu(t)+\gamma
_{t})d\widetilde{W}(t)\\
&  =\frac{1}{2}\frac{\partial^{2}v}{\partial x^{2}}(M_{\gamma}(t))^{2}%
[(\mu(t)+\gamma_{t})^{2}-(\mu(t)+\hat{\gamma}_{t})^{2}]dt+\frac{\partial
v}{\partial x}M_{\gamma}(t)(\mu(t)+\gamma_{t})d\widetilde{W}(t).
\end{align*}
By the definition of $\hat{\gamma}_{t}$ (\ref{hatnu2}), we have $(\mu
(t)+\gamma_{t})^{2}-(\mu(t)+\hat{\gamma}_{t})^{2}\geq0,\ t\in\lbrack0,T]$. The
convexity of $v(t,\cdot)$ guarantees that $v(t,M_{\gamma}(t))$ is a
submartingale. Thus, $\forall\gamma\in\mathcal{B}$,
\[
E_{\gamma}[\tilde{u}(\zeta Z_{\gamma}(T))]=\tilde{E}[g(M_{\gamma}%
(T))]=\tilde{E}v(T,M_{\gamma}(T))\geq\tilde{E}v(0,M_{\gamma}(0))=\tilde
{E}[g(M_{\hat{\gamma}}(T))]=E_{\hat{\gamma}}[\tilde{u}(\zeta Z_{\hat{\gamma}%
}(T))].
\]
This completes the proof. $\ \ \ \ \ \Box$
\end{example}

\begin{example}
Suppose that $|\mu(\cdot)|\leq K$, a.e., a.s.. Then we have
\begin{equation}
\hat{\gamma}_{t}=-\hat{\mu}(t),\ t\in\lbrack0,T],a.s..
\end{equation}
Note that $\mu$ belongs to $\mathcal{B}$ when $|\mu(\cdot)|\leq K$, a.e. a.s..
Due to the convexity of $g$, we have that $\forall\gamma\in\mathcal{B}$,
\[
\tilde{E}[g(M_{\gamma}(t))]\geq g(\tilde{E}(M_{\gamma}(T)))=g(1)\equiv
g(M_{\hat{\gamma}}(T))\equiv\tilde{E}[g(M_{\hat{\gamma}}(T))].
\]
In this case, $P_{\hat{\gamma}}$ coincides with the risk neutral probability
$\widetilde{P}$ on $\mathcal{G}_{T}$ which leads to the optimal terminal
wealth $\hat{\xi}=x$. This means that the investor will not invest on the risk
assets at all.
\end{example}

\section{Terminal perturbation method}

When the generator of the recursive utility (\ref{BSDE}) is non-concave, the
dual method is not applicable. In this case, we apply the terminal
perturbation method to obtain a characterization of the optimal terminal
wealth. We need the following smooth assumption:

\begin{assumption}
\label{assfu1} $f$ is continuously differentiable in $(y, z)$.
\end{assumption}

Let $\xi^{\ast}$ be an optimal terminal wealth for (\ref{optmb}), i.e.
\[
Y^{\xi^{\ast}}(0)=\sup_{\xi\in\mathcal{A}(x)}Y^{\xi}(0),
\]
and $(X^{\ast}(\cdot),q^{\ast}(\cdot),Y^{\ast}(\cdot),Z^{\ast}(\cdot))$ be the
corresponding state processes of (\ref{backsystem}).

Set
\[
\bar{\Omega}:=\{\omega\in\Omega|\xi^{\ast}(\omega)=0\}.
\]
By the terminal perturbation method in \cite{JP} and \cite{JZ}, we have the
following stochastic maximum principle.

\begin{theorem}
Under assumptions \ref{assf1}, \ref{assu1} and \ref{assfu1}, if $\xi^{\ast}$
is the optimal wealth of problem \ref{optmb}, then there exists $h_{0}%
\in\mathbb{R},\ h_{1}\geq0$ and $|h_{0}|^{2}+h_{1}^{2}=1$ such that
\begin{align*}
&  h_{0}m(T)+h_{1}u^{\prime}(\xi^{\ast})n(T)\geq0,\ \ a.s.\ \ on\ \ \bar
{\Omega};\\
&  h_{0}m(T)+h_{1}u^{\prime}(\xi^{\ast})n(T)=0,\ \ a.s.\ \ on\ \ \bar{\Omega
}^{c},
\end{align*}
where
\[%
\begin{cases}
dm(t)=-\hat{\mu}^{\prime}(t)\sigma^{\prime-1}(t)m(t)d\widehat{W}%
(t),\ \ m(0)=1;\\
dn(t)=f_{Y}^{\ast}(t)n(t)dt+f_{Z}^{\ast^{\prime}}(t)n(t)d\widehat{W}%
(t),\ \ n(0)=1,
\end{cases}
\]
and $f_{Y}^{\ast}(t):=f_{Y}(t,Y^{\ast}(t),Z^{\ast}(t))$, $f_{Z}^{\ast
}(t):=f_{Z}(t,Y^{\ast}(t),Z^{\ast}(t))$.
\end{theorem}

\begin{remark}
Note that we do not need the concavity property of $u$ in the above theorem.
\end{remark}

\bigskip\

\renewcommand{\refname}

\end{document}